\documentclass[aip,pop,reprint]{revtex4-1}
\usepackage{amssymb}
\usepackage{amsmath}
\usepackage{amsthm}
\usepackage{nicefrac}
\usepackage{graphicx}
\usepackage{epstopdf}
\usepackage[bookmarks,colorlinks,breaklinks]{hyperref}
\hypersetup{linkcolor=blue,citecolor=blue,filecolor=black,urlcolor=blue}

\makeatletter
\newcommand*{\citenumns}[2][]{%
  \begingroup
  \let\NAT@mbox=\mbox
  \let\@cite\NAT@citenum
  \let\NAT@space\NAT@spacechar
  \let\NAT@super@kern\relax
  \renewcommand\NAT@open{}%
  \renewcommand\NAT@close{}%
  \cite[#1]{#2}%
  \endgroup
}
\makeatother

\begin{document}

\title{Modeling turbulent energy behavior and sudden viscous dissipation \\ in compressing plasma turbulence}

\author{Seth Davidovits}
\affiliation{Princeton University, Princeton, New Jersey 08544, USA}
\author{Nathaniel J. Fisch}
\affiliation{Princeton University, Princeton, New Jersey 08544, USA}
\affiliation{Princeton Plasma Physics Laboratory, Princeton, New Jersey 08544, USA}

\begin{abstract}
We present a simple model for the turbulent kinetic energy behavior of subsonic plasma turbulence undergoing isotropic three-dimensional compression, such as may exist in various inertial confinement fusion experiments or astrophysical settings. The plasma viscosity depends on both the temperature and the ionization state, for which many possible scalings with compression are possible. For example, in an adiabatic compression the temperature scales as $1/L^2$, with $L$ the linear compression ratio, but if thermal energy loss mechanisms are accounted for, the temperature scaling may be weaker. As such, the viscosity has a wide range of net dependencies on the compression. The model presented here, with no parameter changes, agrees well with numerical simulations for a range of these dependencies. This model permits the prediction of the partition of injected energy between thermal and turbulent energy in a compressing plasma.
\end{abstract}

\maketitle

\section{Introduction}\label{sec:introduction_modeling}
When plasma is compressed, any embedded turbulence is also compressed, with the result that the kinetic energy associated with the turbulence may increase. Experimental results in gas-puff Z-pinches suggest that the plasma contains substantial non-radial hydrodynamic motion at stagnation \citep{kroupp2011,maron2013}. The Reynolds number at stagnation is large ($\mathrm{Re} \sim 10^{4} - 10^{5}$), such that this hydrodynamic motion could be turbulent \citep{kroupp2017}. Although the source of the hydrodynamic motion is unclear, it may be carried along (and compressed) during the implosion. Detailed simulations of inertial fusion implosions suggest the presence of turbulence to varying degrees \citep{thomas2012,weber2014}. In the context of astrophysics, turbulence is ubiquitous in interstellar gas \citep{elmegreen2004}, including in molecular clouds, where the impact of gravitational compression on turbulence is studied \citep{robertson2012,davidovits2017}. 

Although there have been a range of modeling efforts for compressing turbulence in neutral gases \citep{morel1982,wu1985,coleman1991,blaisdell1991,speziale1991,durbin1992,cambon1992,coleman1993,blaisdell1996,hamlington2014,grigoriev2016}, the viscosity growth with compression in neutral gas is more restricted than for plasma, necessitating new models capable of capturing plasma phenomena, such as sudden viscous dissipation \citep{davidovits2016a,davidovits2016b}. We present one such new model here. The model aims to capture the turbulent kinetic energy (TKE) behavior of compressing plasma turbulence. To the extent it successfully does so, it can be used to give a partial answer to a key question for plasma turbulence undergoing compression. What is the partition of input energy between thermal energy and TKE for a given compression? This point is addressed in Sec.~\ref{sec:energy_partition_modeling}. A related key question addressed in that section is the total energy required to compress the plasma (total injected energy), given the initial TKE. Section~\ref{sec:energy_partition_modeling} can be read without having gone through the details of the dissipation rate for the turbulence represented by the present model.

The present work, as in much previous work on compressing plasma turbulence or compressing neutral gas turbulence, considers isotropic, three-dimensional (3D), constant velocity compression of homogeneous turbulence. The plasma is modeled as a fluid, but, unlike most previous treatments, with a plasma viscosity. The plasma viscosity depends sensitively on temperature and charge state, $\mu \sim T^{5/2}/Z^4$, with $T$ the temperature and $Z$ the charge state. Either (or both) of $Z$ or $T$ may change during the compression. The amount of compression is indicated by $\bar{L} = L/L_0$, which is the linear compression ratio, so that, for example, $\bar{L} = 10$ corresponds to a reduction in volume by a factor $1/\bar{L}^3 = 1/1000$. Similarly to \citet{davidovits2016b}, we assume that the net effect of changes to $Z$ and $T$ during compression is that the (dynamic) viscosity can be written as a function of $L$,
\begin{equation}
\mu = \mu_0 \bar{L}^{-2 \beta}. \label{eq:viscosity_modeling}
\end{equation}
As a ``base'' plasma case we consider a fixed ionization state ($Z=\mathrm{constant}$, e.g. fully ionized), and a temperature dependence on compression of $T \sim \bar{L}^{-2}$. This is the temperature dependence for adiabatic, 3D, compression of a monatomic ideal gas. With $\mu \sim T^{5/2}/Z^4$, this gives $\mu \sim \bar{L}^{-5}$, or $\beta = 5/2$ for the ``base'' case. When $Z=\mathrm{constant}$, achieving $\beta > 5/2$ would require a stronger temperature growth with compression than $T \sim \bar{L}^{-2}$; this could be achieved by, say, having a heating source in addition to the compression, but still no thermal losses. If changing ionization state is included, so that $Z$ can increase (decrease) during the compression, $\beta$ could be greatly reduced (enhanced). Similarly, if thermal losses are included, $\beta$ will be reduced.

The model given in this work matches reasonably well with simulations over the range $\beta \in [1,5/2]$, with tests at $\beta = 1, 3/2, 5/2$. It may also be valid outside this range, but has not been tested there. In addition to the ``base case'' of a plasma heating adiabatically with fixed ionization state ($\beta = 5/2$), other cases of interest will be included in the validity range.  As an example, \citet{lindl1995} showed that for a hot spot where the temperature is determined by the balance of mechanical heating and thermal conduction, $T \sim \left(\rho \bar{L} \right)^{2/5}$. With $\rho \sim \bar{L}^{-3}$, this means $T \sim \bar{L}^{-4/5}$, which corresponds to the case $\beta = 1$, when $Z$ is constant, as in a deuterium-tritium hot spot with no mix.   

For $\beta < 1$, the present model predicts a saturation of the TKE under continuing compression; see Sec.~\ref{sec:discussion_modeling} for a discussion of the model behavior in this range. In addition to neutral gases, a case of interest falling in the $\beta < 1$ range is isothermal turbulence, where both $T$ and $Z$ are constant, corresponding to $\beta = 0$. Compressing isothermal turbulence is of interest for understanding molecular clouds \citep{robertson2012,davidovits2017}, where supersonic turbulence is compressed by pressure or gravity. The present model applies, strictly speaking, only to subsonic turbulence. 

The model may be valid for $\beta > 5/2$ (there is no reason to expect a sudden transition at this value in the turbulence behavior under compression), but we have not tested it in this regime. The inertial fusion applications we have in mind typically do not have temperature growth with compression strong enough to bring them into this range, once thermal energy loss mechanisms like conduction and radiation are taken into account. Two examples of heating mechanisms that could potentially push the temperature growth with compression into this range are fusion heating, and heating from the sudden dissipation of TKE, if the dissipated TKE is fed back self-consistently into temperature. Note that fusion typically occurs after the compression is essentially over, while the sudden dissipation of TKE, if it is truly sudden, would occur at nearly fixed $\bar{L}$.

The structure of this paper is as follows. Section \ref{sec:equations_modeling} describes the system of equations governing the compressing turbulence. The following section, Sec.~\ref{sec:model_modeling} describes the model for the TKE of the system. Section \ref{sec:comparison} discusses the setting of model constants and shows comparisons between the model and numerical simulations. The way in which the model addresses the partition of input compression energy between thermal energy and TKE is  discussed in Sec.~\ref{sec:energy_partition_modeling}; also discussed is the total energy injected. Finally, Sec.~\ref{sec:discussion_modeling} provides additional discussion and caveats.

\section{Governing equations} \label{sec:equations_modeling}

The system of equations governing the compressing turbulence is the same as that described in \citet{davidovits2016b}, which in turn is very similar to previous work in neutral gases (e.g.~[\citenumns{wu1985,coleman1991,blaisdell1991,cambon1992,hamlington2014}]). There are also similarities with previous work in astrophysics (e.g.~[\citenumns{robertson2012}] and \citet{peebles1980}, Section 9), although here we consider only subsonic turbulence. For a full derivation and discussion of the system, readers are referred to Appendix Sec. 1 in \citet{davidovits2016b}.

The system is the Navier-Stokes (NS) equations, with the flow, $\mathbf{v}$, broken into an imposed background flow and an unknown (turbulent) component, $\mathbf{v} = \mathbf{v}_{0} + \mathbf{v'}$. The imposed background flow, $\mathbf{v}_{0}$, has the form, 
\begin{equation}
 \mathbf{v}_{0} \! \left(\mathbf{x},t\right) = (\dot{L}/L ) \mathbf{x}, \label{eq:background_flow_modeling}
\end{equation}
 with the overdot the time derivative. When $L$ is defined
\begin{equation}
L\! \left( t \right)  = L_0 - 2 U_b t, \label{eq:L_modeling}
\end{equation}
the effect of the background flow is as follows; a homogeneous initial turbulent field remains homogeneous, and a 3D box of initial side length $L_0$, advected by the background flow, will remain a box and have a side length given by $L \left( t \right)$.  That is, the sides of this compressing box move inward at a constant velocity $U_b$. 

The turbulent field itself, $\mathbf{v'}$, is treated in the low-Mach (incompressible) limit. This means density fluctuations can be ignored, and the continuity equation gives the expected density dependence for a 3D compression,
\begin{equation}
\rho_0 \! \left( t \right) = \rho_0 \! \left( 0 \right) \left( L_0 / L\! \left( t \right) \right)^3 = \rho_0 \! \left(0 \right) \bar{L}^{-3}. \label{eq:rho_modeling}
\end{equation}
Here, $\rho_0 ( 0 )$ is the initial (uniform) density.

Under these conditions, and working in a coordinate system that is co-moving with the background flow, the NS momentum equation for the turbulence is,
\begin{equation}
\frac{\partial \mathbf{V}}{\partial t}+\frac{1}{\bar{L}}\mathbf{V}\cdot \nabla \mathbf{V}-\frac{2U_{b}}{L}\mathbf{V}+\frac{\bar{L}^2}{\rho_{0}\!\left(0\right)} \nabla P  =  \nu_0 \bar{L}^{1-2\beta}\nabla^2 \mathbf{V}. \label{eq:moving_momentum_modeling}
\end{equation}
The third term is a forcing term associated with the compression. The initial kinematic viscosity is $\nu_0 = \mu_0/\rho_0 (0)$, and Eq.~(\ref{eq:viscosity_modeling}) has been used for the viscosity dependence on $\bar{L}$. The turbulent velocity field $\mathbf{V}$ is simply $\mathbf{v}'$ rewritten in the moving coordinates, $\mathbf{V} (\mathbf{X},t) = \mathbf{v}' (\mathbf{x},t )$, where $\mathbf{x} = \bar{L} \mathbf{X}$. This means that the TKE behavior determined in the moving coordinates is the same as the laboratory frame TKE behavior. We consider Eq.~(\ref{eq:moving_momentum_modeling}) on a cubic domain with sides extending from $-L_0/2$ to $L_0/2$, and periodic boundary conditions.

The (ensemble averaged) TKE, $E$, is defined as,
\begin{equation}
E = \langle \mathbf{V}^2 /2 \rangle, \label{eq:mean_TKE}
\end{equation}
with the angle brackets indicating the average, which in this homogeneous case can be taken to be a spatial average. The total energy in the domain (excluding the background flow) is then $E_{\rm{tot}} = \rho_0 (0) \langle \mathbf{V}^2 \rangle L_0^3/2$. 

\begin{figure*}
\includegraphics{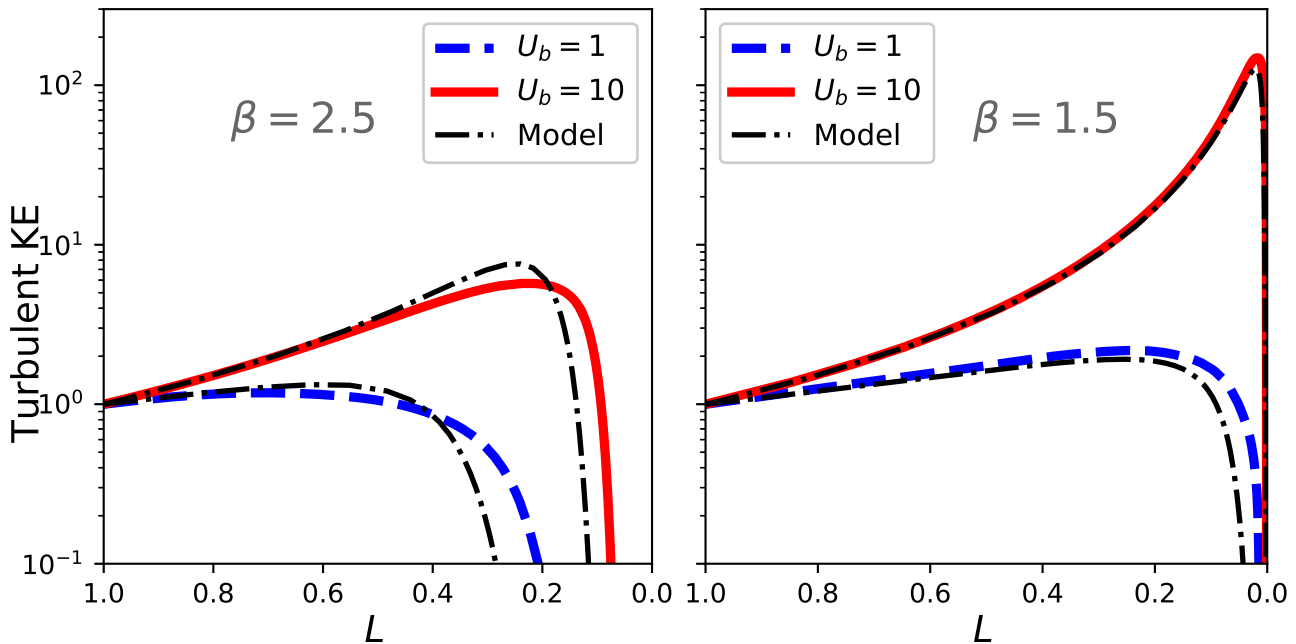}
\caption{A comparison of simulation and model turbulent kinetic energy (TKE) during compression at two rates, for two different net viscosity dependencies on compression, $\mu \left(\bar{L} \right) \sim \bar{L}^{-2 \beta}$; see Eq.~(\ref{eq:viscosity_modeling}) and the surrounding discussion. On the left, $\beta=2.5$, the ``base'' plasma case with adiabatic plasma heating and fixed ionization state. On the right, $\beta =1.5$, representing either weaker heating, and/or some ionization during compression. An initially turbulent flow (Taylor-Reynolds number $\rm{Re}_{\lambda,0} \approx 82$) is compressed at a velocity $U_b$ on times equal to ($U_{b,\mathrm{norm}}=1$) or faster than ($U_{b,\mathrm{norm}} = 10$) the initial turbulent decay time. The x-axes are the linear compression ratio; the domain is a box of initial side length 1, with the side length shrinking as the compression progresses from left to right in the graphs. For each simulation, we also plot the result of the model with the same $U_b$, $\beta$, $\nu_0$ and initial TKE, Eq.~(\ref{eq:scaled_model_L}), rescaled to the lab frame using Eq.~(\ref{eq:energy_transformation}). The results show reasonable agreement, with no ``free'' parameters used between the two different $\beta$ cases. There is an apparent tendency for the model to suddenly dissipate somewhat early (at larger $L$). The simulation and model parameters are given in full in Table \ref{tbl:parameters}.
\label{fig:twoBetas}}
\end{figure*}

\begin{figure}
\includegraphics{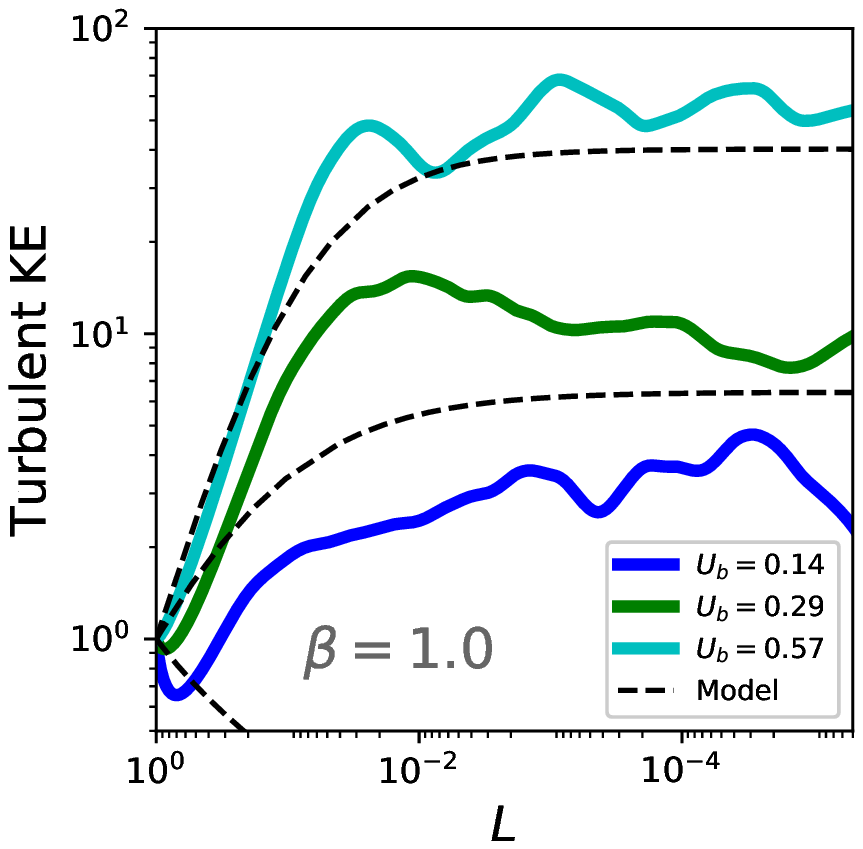}
\caption{Same as Fig.~\ref{fig:twoBetas}, but for $\beta=1.0$, at a lower initial Reynolds number ($\rm{Re}_{\lambda,0} \approx 14$), and with a logarithmic scale for $L$. A low initial Taylor-Reynolds number and modest compression speeds are used to ensure the turbulence is resolved at saturation. The model correctly predicts that the turbulent kinetic energy saturates, although at a level below the actual saturation for these cases. This difference will disappear at higher initial Reynolds numbers or for faster compressions. One can observe that the saturation level is closer to the true value for $U_{b,\mathrm{norm}} = 0.57$ than for $U_{b,\mathrm{norm}} = 0.29$. At the slowest compression speed, $U_{b,\mathrm{norm}} = 0.14$, the model predicts that the energy decays, while it actually saturates at a finite level. The simulation in this case is near the boundary in parameter space where the TKE decays under compression; the model has the location of this boundary somewhat wrong. For more discussion of the $\beta = 1$ case, see Sec.~\ref{sec:comparison}. The simulation and model parameters are given in full in Table \ref{tbl:parameters}.
\label{fig:beta1}}
\end{figure}

\section{TKE model} \label{sec:model_modeling}

We seek a model that will give the TKE behavior of the compressing turbulence described by Eq.~(\ref{eq:moving_momentum_modeling}).  By rescaling $\mathbf{V}, P, t$ by appropriate powers of $\bar{L}$, this equation can be rewritten as the usual, unforced, NS equation, but with a time-varying viscosity. In the rescaled equation the turbulence will decay. Given a model for the TKE behavior during this decay, one can translate the model back from the rescaled variables, yielding a model for the compressing turbulence TKE behavior. This observation has been used by \citet{cambon1992} to aid modeling efforts for compressing turbulence previously, but with the viscosity variation assumed to be negligible. Here, we expand on this work by including the viscosity variation and low Reynolds number effects. These effects are found to have a substantial impact on the TKE behavior in the regime of interest.

Using the rescalings,
\begin{eqnarray}
\mathbf{V} & = & \bar{L}^{-1}\hat{\mathbf{V}},\\
P & = & \bar{L}^{-5}\hat{P},\\
\rm{d}\hat{t} & = & \bar{L}^{-2}\rm{d}t, \label{eq:time_transformation}
\end{eqnarray}
in Eq.~(\ref{eq:moving_momentum_modeling}) gives,
\begin{equation}
\frac{\partial\hat{\mathbf{V}}}{\partial\hat{t}}+\hat{\mathbf{V}}\cdot \nabla \hat{\mathbf{V}} =
- \frac{1}{\rho_0 \! (0)}\nabla \hat{P}+ \nu_0 \bar{L}^{3 - 2\beta} \nabla^2 \hat{\mathbf{V}}.\label{eq:scaled_momentum_modeling}
\end{equation}
The rescaled TKE is $\hat{E} = \langle \hat{V}^2/2 \rangle$, and it will be related to the laboratory frame TKE by,
\begin{equation}
E = \langle \frac{\mathbf{V}^2}{2} \rangle = \bar{L}^{-2} \langle \frac{\hat{\mathbf{V}}^2}{2} \rangle = \bar{L}^{-2} \hat{E}. \label{eq:energy_transformation}
\end{equation}
Once again, the angle brackets indicate an ensemble average, which in this homogeneous case can be taken to be a spatial average. When the viscosity coefficient in Eq.~(\ref{eq:scaled_momentum_modeling}) has no time dependence ($\beta = 3/2$), this scaled equation is the usual NS momentum equation. At high Reynolds numbers the decay of NS turbulence closely follows a power law in time that is independent of the viscosity, $\hat{E} \sim \hat{t}^{-n}$ (see \citet{sinhuber2015} and references therein). Neglecting viscosity, the power $n$ depends on whether the energy containing turbulent scale is free to grow ($n \sim 1.2$) or not ($n \sim 2$) \citep{skrbek2000}. One could translate this high Reynolds number expression for $\hat{E} (\hat{t})$ into one for the lab-frame energy $E (t)$, using Eqs.~(\ref{eq:time_transformation},\ref{eq:energy_transformation}). However, using this power law decay to capture the behavior of $\hat{E}$ neglects viscous effects; the decay is determined solely by the rate of energy cascade from the large scales. It has been shown, however, that varying only $\beta$, and therefore only the viscous coefficient time dependence, can make a huge difference in the TKE behavior of compressing turbulence (see \citet{davidovits2016b}, in particular Fig.~3, see also \citet{coleman1991}). Thus, to model this TKE behavior, we must account somehow for the viscous effects.

The present approach to do this accounting is to use the high to low Reynolds number decay model of \citet{lohse1994}, with the additional step of assuming the viscosity in that model is time dependent. This assumption is convenient, rather than rigorous, but yields reasonable results for the cases examined. Lohse's model for the energy behavior of the decaying NS equation Eq.~(\ref{eq:scaled_momentum_modeling}), but with time dependent viscosity, is
\begin{equation}
 \frac{\left(12/3b\right)^{3/2}}{L_{\rm{outer}}} \left[ \nu_n \! \left( \hat{t} \right) + \sqrt{\hat{E}\! \left( \hat{t} \right) + \left( \nu_n \! \left( \hat{t} \right) \right)^2} \right]  \hat{E} \! \left( \hat{t} \right) =  - \frac{\rm{d} \hat{E}\! \left( \hat{t} \right)}{\rm{d} \hat{t}} . \label{eq:scaled_model}
\end{equation}
Both terms on the left hand side of Eq.~(\ref{eq:scaled_model}) cause $\hat{E}(\hat{t})$ to decrease in $\hat{t}$; the first term within the square braces comes from viscous dissipation, while the second (square root) term is a combination of viscous dissipation ($\nu$ part) and an effective eddy viscosity ($\hat{E}$ part). The model is completed with the definition,
\begin{equation}
\nu_n \! \left( \hat{t} \right)  =  \frac{3 }{4} b^{3/2} \frac{\nu_0}{L_{\rm{outer}}} \frac{\nu \! \left( \hat{t} \right)}{\nu_0} = B \frac{\nu \! \left( \hat{t} \right)}{\nu_0},
\end{equation}
so that
\begin{equation}
B = \frac{3 }{4} b^{3/2} \frac{\nu_0}{L_{\rm{outer}}}. \label{eq:B_definition}
\end{equation}
Here $b$ is the Kolmogorov constant, and $L_{\rm{outer}}$ is a (constant) length scale associated with the large scales of the turbulence (as in \citet{lohse1994}). Lohse's model is recovered by setting $\nu_n (\hat{t}) = B$. It is convenient to work in terms of $\bar{L}$, rather than $\hat{t}$. This is achieved by making use of Eqs.~(\ref{eq:L_modeling},\ref{eq:time_transformation}). Doing so gives,
\begin{equation}
A \left[ B \bar{L}^{3-2\beta} + \sqrt{\hat{E}\! \left( \bar{L} \right) + \left( B \bar{L}^{3-2\beta} \right)^2} \right]  \frac{\hat{E} \! \left( \bar{L} \right)}{\bar{L}^2} =   \frac{\rm{d} \hat{E}\! \left( \bar{L} \right)}{\rm{d} \bar{L}} . \label{eq:scaled_model_L}
\end{equation}
We have also substituted for the appropriate viscosity, which is the coefficient of the last term in Eq.~(\ref{eq:scaled_momentum_modeling}), $\nu = \nu_0 \bar{L}^{3 - 2 \beta}$. The constant A is defined,
\begin{equation}
A = 8 b^{-3/2} \frac{L_0}{2 U_b} \frac{1}{L_{\rm{outer}}}. \label{eq:A_definition}
\end{equation}
The domain of the model, Eq.~(\ref{eq:scaled_model_L}), is $\bar{L} \in \left(0,1 \right]$, with the initial condition (the starting energy) being set at $\bar{L} = 1$.
After setting values of the constants $b$ and $L_{\rm{outer}}$, the model, Eq.~(\ref{eq:scaled_model_L}) can be solved for $\hat{E}$ for the set of physical parameters, $\left\{\beta,U_b,\nu_0,L_0 \right\}$, of interest. Having solved for $\hat{E}  (\bar{L})$, the untransformed energy is obtained by rescaling by $\bar{L}^{-2}$, as per Eq.~(\ref{eq:energy_transformation}).

\begin{table}
    \caption{\label{tbl:parameters} Parameters for each simulation plotted in Figs. \ref{fig:twoBetas} and \ref{fig:beta1}, and the corresponding calculated model coefficients (rounded to two significant figures). The domain for all cases is a periodic cube with initial side length $L_0 = 1$. The initial TKE in the simulations is $E_0$, Eq.~(\ref{eq:mean_TKE}) evaluated at $\bar{L}=1$ ($t=0$). Since $\bar{L}=1$ initially, $E_0 = \hat{E}_0$, so that the initial scaled TKE is the same as the initial lab frame energy; the two are related by Eq.~(\ref{eq:energy_transformation}). The nonlinearity of the model, Eq.~(\ref{eq:scaled_model_L}), with respect to the energy, $\hat{E}$, means that the precise initial value matters. After solving for $\hat{E} (\bar{L})$, it is rescaled for plotting in Figs. \ref{fig:twoBetas} and \ref{fig:beta1} by dividing by $E_0$ ($\hat{E}_0$). The initial mean viscous dissipation, $\epsilon_0$, is $\epsilon = -\nu_0 \langle \mathbf{v} \cdot \nabla^2 \mathbf{v} \rangle$ evaluated for the initial flow field (at $\bar{L} = 1$). Together with $E_0$, it sets the initial turnover time, $\tau_t = E_0/\epsilon_0$, which in turn determines the $U_b$ needed to achieve the desired normalized compression rate $U_{b,\mathrm{norm}}$, which is shown in Figs. \ref{fig:twoBetas} and \ref{fig:beta1}. Since $L$ is defined by Eq.~(\ref{eq:L_modeling}), the compression time is $\tau_c = L_0/2 U_b$, and we can find that $U_b = (L_0 / 2 \tau_t) U_{b,\mathrm{norm}}$ will give a compression where $\tau_t/\tau_c = U_{b,\mathrm{norm}}$. Also shown are a (derived) initial Taylor-Reynolds number for each initial state, $\mathrm{Re}_{\lambda,0} = (2 E_0 \sqrt{15 \nu_0/\epsilon_0})/\nu_0$. The definitions of $B$ and $A$ for the TKE model are given in Eqs.~(\ref{eq:B_definition}) and (\ref{eq:A_definition}).
    All model cases use $b \approx 5.949$ and $L_{\mathrm{outer}} \approx 0.19 L_0$, as described in Sec.~\ref{sec:comparison}.
    }
    \begin{tabular}{cc|ccccc|ccc}
    \hline
    \multicolumn{2}{c|}{Case} &
    \multicolumn{5}{c|}{Simulation} &
    \multicolumn{3}{c}{Model}\\
    $\beta$ & $U_{b,\mathrm{norm}}$ & 
    $\nu_0$ & $E_0$ & $\epsilon_0$ & $U_b$ & $\mathrm{R}_{\lambda,0}$  
    & $A$ & $B$ & $\hat{E}_0$  \\
    \hline
    2.5 & 1 & 1/600 & 0.34 & 0.62 & 0.91 & 82 & 1.6 & 0.095 & 0.34  \\
    2.5 & 10 & 1/600 & 0.34 & 0.62 & 9.1 & 82 & 0.16 & 0.095 & 0.34  \\
    1.5 & 1 & 1/600 & 0.34 & 0.62 & 0.91 & 82 & 1.6 & 0.095 & 0.34  \\
    1.5 & 10 & 1/600 & 0.34 & 0.62 & 9.1 & 82 & 0.16 & 0.095 & 0.34  \\
    1.0 & 0.14 & 1/100 & 0.34 & 3.7 & 0.78 & 14 & 1.9 & 0.57 & 0.34  \\
    1.0 & 0.29 & 1/100 & 0.34 & 3.7 & 1.6 & 14 & 0.93 & 0.57 & 0.34  \\
    1.0 & 0.57 & 1/100 & 0.34 & 3.7 & 3.1 & 14 & 0.46 & 0.57 & 0.34  \\
    \hline
    \end{tabular}
\end{table}

\section{Setting constants and simulation comparison} \label{sec:comparison}
Here we compare the solutions of the model, Eq.~(\ref{eq:scaled_model_L}), to simulations of compressing turbulence as described by Eq.~(\ref{eq:moving_momentum_modeling}). To set values of $b$ and $L_{\rm{outer}}$, we use the following procedure. When $\beta = 1$, compressing turbulence described by Eq.~(\ref{eq:moving_momentum_modeling}) will reach a steady state energy as $\bar{L} \rightarrow 0$ (see \citet{davidovits2016b}). This steady state energy is $E = 1.9 U_b^2$. The characteristic large scale associated with the steady state is $L_{\rm{outer}} \approx 0.19 L_0$ \citep{rosales2005,davidovits2016b}. We choose this to be the value of $L_{\rm{outer}}$. In the limit $L \rightarrow 0$, when $\beta =1$, the model, Eq.~(\ref{eq:scaled_model_L}), also predicts that $E$ reaches a steady state (that is, it predicts $\hat{E} \sim \bar{L}^2$). The model gives for $E$,
\begin{equation}
E\! \left( \bar{L} \rightarrow 0 \right) = \bar{L}^{-2} \hat{E} \! \left( \bar{L} \rightarrow 0 \right) = \frac{4}{A^2} \left(1 - B A \right). \label{eq:steady_energy}
\end{equation}
Setting $b \approx 5.949$ makes $4/A^2 \approx 1.9 U_b^2$. Then the leading term in Eq.~(\ref{eq:steady_energy}) matches the expected result. Further, this value of $b$ is not far off from experimental values of the Kolmogorov constant (see \citet{lohse1994} for more discussion on values of $b$). The term $B A$ itself has no dependence on $b$; this term can be rewritten in terms of the ratio $\tau_c/\tau_\nu$ of the compression timescale, $\tau_c = L_0/2 U_b$, to the viscous timescale, $\tau_\nu = L_{\rm{outer}}^2/\nu_0$. For rapid compressions, or compressions with small initial viscosity, this will be very small, so that $1 - B A \approx 1$, and the energy saturation predicted by the model will match the expected result.

Figures \ref{fig:twoBetas} and \ref{fig:beta1} show comparisons between the model with $b$ and $L_{\rm{outer}}$ set as just described, and direct numerical simulations of compressing turbulence carried out in the spectral code Dedalus \cite{dedalus}. For the initial condition we generate a turbulent flow field using Lundgren's method \cite{lundgren2003,rosales2005}. This turbulence is then compressed, evolving according to Eq.~(\ref{eq:moving_momentum_modeling}) (the simulations actually use a rescaled version of Eq.~(\ref{eq:moving_momentum_modeling}), and then the results are appropriately scaled back, see [\citenumns{davidovits2016a,davidovits2016b}]). All simulations use a $192^3$ Fourier grid, dealiased to $128^3$, and periodic boundary conditions. For each comparison, $U_b, E (\bar{L} = 1), \nu_0, \beta$, and $L_0$ are set in the model to the simulation values. Thus, having set $b$ and $L_{\rm{outer}}$ once, there are no ``free'' parameters used in the comparisons. Table \ref{tbl:parameters} gives quantities that describe the initial turbulent state that undergoes compression in each simulation presented in Figs. \ref{fig:twoBetas} and \ref{fig:beta1}. The table also gives the values for $U_b, E (\bar{L} = 1) = E_0, \nu_0, \beta$, and the values of $A$ and $B$ for the matching model.

The figure captions contain more discussion of the comparison. Here we comment on the disagreement between the model and simulations for the $\beta = 1, U_b = 0.14$ case in Fig.~\ref{fig:beta1}. One can show that, if the ratio $U_b/\nu_0$ is too small, the energy will purely decay in the $\beta = 1$ case, rather than reach the saturated state (\citet{davidovits2016b}). The model predicts a similar breakdown, if $1 - B A < 0$. However, with the present choices of $b$ and $L_{\rm{outer}}$, the value of the ratio $U_b/\nu_0$ for which pure decay occurs in the model is not quite correct. It can be made correct, while simultaneously keeping the agreement with the $\beta = 1$ steady state energy, with a different choice for $b$ and $L_{\rm{outer}}$. We find this requires a value for $b$ far outside the reasonable range, and, while reducing the error for the $\beta = 1$ case, increases the error in the other cases. Note that the simulations in Fig.~\ref{fig:beta1} are carried out with relatively high initial viscosity (low Reynolds number) to keep them resolved at saturation for our modest resolution. It is only at these low initial Reynolds numbers and for slow compressions that the pure decay occurs (in either the model or simulations).

\section{Energy partition and total injected energy} \label{sec:energy_partition_modeling}

\subsection{Coupled model, energy partition}

By coupling the model developed here to a temperature equation, we can examine how the energy injected by the compression is partitioned between thermal energy and turbulent energy. Although the model has not been tested against simulations with this type of coupling (a fact which is discussed further at the end of this subsection), we present this coupled system to place the modeling efforts of this paper in context, and to make more clear the end goal and some of the outstanding problems. For this sake, the origin of the present model for the turbulent dissipation, $\bar{\epsilon}_{\rm{model}}$, is not important.

To write a coupled system for the temperature and TKE, for the case of the plasma viscosity, we revise Eq.~(\ref{eq:viscosity_modeling}). Instead of assuming $T \sim \bar{L}^{-2}$, we treat $T$ as an unknown, and use the plasma viscosity dependence on temperature, $\mu = \mu_0 T^{5/2}$. At present, we assume $Z = 1$, although it would be straightforward to include a (given) $Z (T)$. Making this substitution of a temperature dependent viscosity, with $T$ unknown, yields for the TKE model equation, instead of Eq.~(\ref{eq:scaled_model_L}),
\begin{equation}
\frac{\rm{d}}{\rm{d} \bar{L}} \bar{E} \! \left( \bar{L} \right) = -2 \frac{\bar{E}}{\bar{L}} + \bar{L}^{-4} \bar{\epsilon}_{\rm{model}},\label{eq:coupled_TKE_modeling}
\end{equation}
where the model turbulent dissipation, $\bar{\epsilon}_{\rm{model}}$, is
\begin{multline}
\bar{\epsilon}_{\rm{model}} \! \left(\bar{L} \right) \equiv \\
2 \frac{\tau_c}{\tau_{\nu}} \left\{ \bar{T}^{5/2}\bar{L}^3 + \sqrt{\frac{1}{2}\frac{V_0^2}{\gamma^2}\bar{L}^2 \bar{E} \! \left( \bar{L} \right) + \bar{T}^5 \bar{L}^6} \right\} \bar{L}^2 \bar{E} \! \left(\bar{L} \right). \label{eq:epsilon_modeling}
\end{multline}
The first term to the right of the equals sign in Eq.~(\ref{eq:coupled_TKE_modeling}) gives ``adiabatic heating'' of the TKE due to the compression, while the second term, which is the dissipation of the TKE, reduces the TKE growth.
At present we work with a normalized laboratory TKE, $\bar{E}$, which has been normalized using an initial turbulent velocity, $V_0$. It can be related to the laboratory TKE, $E$, as
\begin{equation}
E \! \left( \bar{L} \right) = \frac{V_0^2}{2} \bar{E} \! \left( \bar{L} \right). \label{eq:normalized_TKE_modeling} 
\end{equation}
Because we have normalized the TKE, it is convenient to convert $A$ and $B$ to a new set of constants. These constants are
\begin{align}
\tau_c &= -\frac{1}{\dot{\bar{L}}} = \frac{L_0}{2 U_b}, \label{eq:compression_time_modeling} \\
\tau_\nu &= \frac{\alpha^2 L_0^2}{3 \nu_0}, \label{eq:viscous_time_modeling} \\
\gamma &= \left( \frac{9 b^3}{16} \right)^{3/2} \frac{\nu_0}{\alpha L_0}. \label{eq:gamma_modeling}
\end{align}
Here $\tau_c$ is the compression time, $\tau_\nu$ is an initial viscous time, and $\gamma$ can be thought of as a ``viscous velocity''. We have defined $L_{\mathrm{outer}} = \alpha L_0$, so that $\alpha \approx 0.19$, using the strategy for picking $L_{\mathrm{outer}}$ outlined in Sec.~\ref{sec:comparison}. Note that $\tau_c$ and $\tau_\nu$ only enter the model as a ratio (this is why we have gone from 2 constants to 3). Also note that, since $\bar{E}$ is now a normalized energy, the initial characteristic turbulent velocity $V_0$ now appears explicitly in the model equation.

This model TKE equation, Eq.~(\ref{eq:coupled_TKE_modeling}), must then be coupled to a temperature equation in order to look at the question of energy partition. The simplest consistent coupled system includes only the mechanical compression and the viscous dissipation in the temperature equation, taking $\bar{T}$ to be,
\begin{equation}
\frac{\rm{d}\bar{T}}{\rm{d}\bar{L}} = -2 \frac{\bar{T}}{\bar{L}} - \mathcal{E}_{\rm{r}0} \bar{L}^{-4}\bar{\epsilon}_{\rm{model}}. \label{eq:T_modeling}
\end{equation}
The first term to the right of the equals sign in Eq.~(\ref{eq:T_modeling}) is the adiabatic heating due to the compression, while the second term is the increase in thermal energy due to dissipated TKE.
The temperature is normalized to the initial temperature, $T_0$, and $\mathcal{E}_{\rm{r}0}$ is the initial ratio of the energy density of TKE to thermal energy,
\begin{equation}
\mathcal{E}_{\rm{r}0} = \frac{E^T_0}{E^{\rm{th}}_0} = \frac{\rho_0 V_0^2}{2}\frac{1}{3 n_0 k_B T_0}. \label{eq:energy_ratio_modeling}
\end{equation}
The thermal energy is written assuming a plasma with $Z=1$, and assuming equal temperature ions and electrons. As noted, the $Z=1$ assumption can be relaxed in principle, and a possibly changing ionization state included in the viscosity used in the model.

The coupled system, Eqs.~(\ref{eq:coupled_TKE_modeling}),~(\ref{eq:T_modeling}), can be used to predict the proportion of thermal energy and turbulent energy after any amount of compression at a given rate. This represents one key component to understanding the behavior of plasma turbulence under compression. Because the present model is, strictly speaking, only for subsonic compressions, it may not be accurate if $\mathcal{E}_{\rm{r}0}$ is too large. 

Note that, if the solution for $T$ resulting from this coupled system is very different from a power law, this is stretching the model into a situation in which it has not explicitly been tested; the model was only compared to simulations where the viscosity (and therefore the underlying temperature) varied as a power law (with constant power). If $\mathcal{E}_{\rm{r}0} \ll 1$, or if the dissipation of the TKE to thermal energy is gradual, then the behavior of $T$ will be primarily driven by the compressive heating, which does result in a power law temperature dependence ($T \sim \bar{L}^{-2}$).

\subsection{Total injected energy}

While the partition between turbulent energy and thermal energy will depend on the specifics of the model for the turbulent dissipation, $\bar{\epsilon}_{\rm{model}}$, we can also say some things that will not depend on the particular functional form of $\bar{\epsilon}_{\rm{model}}$. Of note is that, for the coupled system, the total energy density, which is proportional to $\mathcal{E}_{\rm{r}0} \bar{E} + \bar{T}$, grows as $\bar{L}^{-2}$ from its initial value of $1+\mathcal{E}_{\rm{r}0}$, 
\begin{equation}
\frac{\rm{d}}{\rm{d}\bar{L}} \left( \mathcal{E}_{\rm{r}0} \bar{E} + \bar{T} \right) = - \frac{2}{\bar{L}} \left( \mathcal{E}_{\rm{r}0} \bar{E} + \bar{T} \right).
\end{equation}
This means that, while the thermal energy and the turbulent energy are not individually state functions of the compression, the \emph{total} energy is a state function, depending only on the initial value and the amount of compression. In the event that loss mechanisms, such as radiation, are added to the system, the total energy will also cease to be a state function of the compression. However, one could imagine that, even in the lossless case presently examined, one might have found that the total energy injected by the compression is not a state function.

The fact that the total energy for the lossless system is a state function is apparently due to a combination of the manner of compression, and the fact that the compression is isotropic in three-dimensions. \citet{davidovits2016b}, particularly in Appendix Sec.~1, describe the compression technique in detail. The compression is caused by a background flow. This compression gives rise to the forcing term in Eq.~(\ref{eq:moving_momentum_modeling}), $2 U_b \mathbf{V}/L$. Dotting the momentum equation with $\mathbf{V}$ to obtain the energy equation, gives an energy forcing term, $2 U_b \mathbf{V}^2/L$. This energy forcing term, with proper normalizations, yields the first term in Eq.~(\ref{eq:coupled_TKE_modeling}). Its dependence is only on the total TKE in the system, a property it shares with the temperature under mechanical compression. This property then carries over to the total energy in the coupled system.

If one assumes a compression-generating background flow that is not isotropic (i.e. instead of $\mathbf{v}_0$ given as in Eq.~(\ref{eq:background_flow_modeling}), taking $v_{0,i} = A_{ij} x_j$, with $A_{ij}$ diagonal but having unequal entries), this state-function property for the total energy can be lost. In the case of two-dimensional compression (letting $A_{33} = 0$, $A_{11}=A_{22} = \dot{L}/L$), the behavior of the total energy can be written,
\begin{equation}
\frac{\rm{d}}{\rm{d}\bar{L}} \left( \mathcal{E}_{\rm{r}0} \bar{E} + \bar{T} \right) = - \frac{4}{3} \frac{1}{\bar{L}} \left( \mathcal{E}_{\rm{r}0} \bar{E} \left[ \frac{3}{2} \frac{\langle V_{\parallel}^2 \rangle}{\langle V^2 \rangle} \right] + \bar{T} \right).
\end{equation}
Here, $\langle V_{\parallel}^2 \rangle  = \langle V_x^2 + V_y^2 \rangle$, if the 2D compression is along the $x$ and $y$ directions. If the turbulent energy is split evenly between the three velocity components, then $\langle V_{\parallel}^2 \rangle/\langle V^2 \rangle = 2/3$, and the total energy will behave as a state function, growing as $\bar{L}^{-4/3}$. However, there is no need for the energy to be split in such an equilibrium manner during a compression, and the amount of total energy change will, in general, depend on the split. The 2D case and this split is not addressed in the present model.

Note that the total energy considered here neglects the (constant) energy associated with the background flow, Eq.~(\ref{eq:background_flow_modeling}). This energy, calculated for the domain considered here and in Refs.~[\citenumns{davidovits2016a,davidovits2016b}] (with initial side length $L_0$), is $E_{\mathrm{bg}} = \rho_0 U_b^2 L_0^3 / 2$.  For fast compressions in the subsonic picture, the energy in the background flow is necessarily substantial compared to the turbulent energy. While one expects that the supersonic case will be similar in the mean, the density -- background velocity correlation could instantaneously play a role, since density perturbations in the supersonic case can be substantial. In real cases, the ``background'' flow may itself convert, in some partition, into thermal energy and turbulent motion as stagnation is approached. For example, gas-puff Z-pinch experiments have found that, at stagnation, the energy from the radially directed (compressing) flow has been converted largely to hydrodynamic motion \cite{kroupp2011,maron2013,kroupp2017}. By enforcing the background flow, the present model will not capture this stagnation process; rather it should apply for the non-radial portion of flow in the compression phase (in 3D compressions).

\section{Discussion} \label{sec:discussion_modeling}

With fixed values of the constants $b$ and $L_{\rm{outer}}$, the model shows reasonable agreement with simulations over a range of compression speeds and viscosity dependencies on compression (values of $\beta$). At the same time, it is a relatively simple model to calculate, involving the solution of a single differential equation, Eq.~(\ref{eq:scaled_model_L}). It is in some respects similar in spirit to two or three (differential) equation $k-\epsilon$ or $k-\epsilon-\tau$ models, some of which have been developed for compressing fluid turbulence (e.g.~[\citenumns{wu1985,coleman1991}]). Such models typically include multiple constants that are determined by fitting to simulations. Straightforward applications of these existing models to the present cases show unsatisfactory results (the large viscosity change with compression for the present cases falls outside of those intended to be treated by, e.g.~[\citenumns{coleman1991}]).

For $\beta < 1$, the model predicts that the TKE saturates under continuing compression, at $E (\bar{L} \rightarrow 0 ) = 4/A^2 = 1.9 U_b^2$. The final equality holds for the choice of $b$ and $L_{\rm{outer}}$ used here. It is unclear whether this should be the case. \citet{davidovits2016b} showed that for $\beta = 1$, the number of linearly forced modes (in Fourier space) is a constant in time (equivalently in $\bar{L}$), in which case the TKE reaching a steady state is unsurprising. When $\beta > 1$, the number of linearly forced modes always eventually reaches zero, so that all modes are damped beyond some time, and thus the TKE will eventually decrease. If $\beta < 1$, the number of linearly forced modes grows indefinitely in time. Naively, one could then expect the TKE to grow indefinitely, disagreeing with the model prediction. However, the compression forces each mode proportional to its TKE content. Assuming a TKE spectrum that decreases monotonically with increasing mode number (decreasing wavelength), each additional forced mode contributes less forcing than the previous one, perhaps yielding a finite total forcing, and a TKE that reaches a steady state under continuing compression. This possibility is hinted at by the fact that the saturated energy in the $\beta = 1$ case does not depend on viscosity, but the number of linearly forced modes does. Thus, two cases with the same compression velocity, but different viscosities, reach the same steady state energy ($E = 1.9 U_b^2$) while having different (but constant) numbers of linearly forced modes.

Since $\beta < 1$ contains some cases of interest, particularly compressing isothermal turbulence and compressing neutral gas turbulence, the question of turbulence saturation (or not) in this regime should be looked at in future work. Whether or not the turbulence saturates may depend on the boundary conditions, and in particular, whether the turbulent length scale saturates or is free to grow.

Compression of the type considered here will cause the turbulent length scale, $L_{\rm{outer}}$, to tend to grow (in the compressing frame). This can be seen from the scaled equation in the moving frame, Eq.~(\ref{eq:scaled_momentum_modeling}); the turbulence decays in these variables, and the length scale in decaying turbulence tends to grow (see, e.g.,~[\citenumns{skrbek2000,sinhuber2015}], or, for the supersonic case, [\citenumns{maclow1999}]). Although the turbulent length scale tends to grow, it can be inhibited by the boundary conditions from doing so --- in the periodic simulations here the length scale is limited by the box size. In a real situation, a physical boundary can limit the maximum length scale.

In Lohse's model, the outer turbulent length scale, $L_{\rm{outer}}$ is fixed, not free to grow. This is reflected in the zero-viscosity-limit decay given by the model, $\hat{E} \sim \hat{t}^{-2}$. One could imagine compressing turbulence with an outer length scale that is either fixed or free to grow, depending on the initial (turbulent) length scale, time of compression, and means of compression. For example, in a capsule-driven compression, the largest (outer) scale possible for the turbulence will be fixed by the capsule size. This will be a constant in the compressing frame. Then, if the initial value of $L_{\rm{outer}}$ is near the capsule size, $L_{\rm{outer}}$ can saturate early in the compression. However, if the initial value of $L_{\rm{outer}}$ is much smaller than the capsule size, then the turbulent length scale may not grow to saturation during the period of compression. For compressions during which the outer length scale grows substantially, the present model may not predict the results well. 

On a related note, we can anticipate that the present model, and its agreement with the simulations, may be improved upon by introducing a more accurate accounting for the value of $L_{\rm{outer}}$. While the turbulent length scale saturates (and becomes a constant under time average) at the value $L_{\rm{outer}} = 0.19 L_0$ in the $\beta = 1$ simulations, $L_{\rm{outer}}$ can neither generally be expected to take this value, nor can it be expected to always be fixed.

These considerations surrounding a fixed or free outer length scale should generally be considered when modeling or simulating turbulence undergoing compression. We speculate that they underlie an apparent disagreement between two results on the behavior of compressing, isothermal, supersonic turbulence. \citet{robertson2012} give a model for the TKE behavior of compressing, isothermal, supersonic turbulence (as a function of compression, similar in spirit to the model here, although with no viscosity variation). With some caveats, \citet{davidovits2017} give a lower bound on this same TKE behavior; that is, show that the TKE of compressing, isothermal, supersonic turbulence must be at least as great as the bound (which is a function of compression). As for the compressing subsonic turbulence treated in the present case, compressing supersonic turbulence can be rescaled into decaying turbulence (with certain restrictions). \citet{davidovits2017} show that the effective decay rate for the model of \citet{robertson2012} must be at least $t^{-2}$, in order for the model to not violate the bound. Since decay rates this large are associated with a fixed outer turbulent length scale, this suggests that that model is, like the present model, most appropriate when the length scale of the compressing turbulence is saturated. However, in the case of astrophysical molecular clouds (the application for the model in \citep{robertson2012}), it is less clear whether a saturated outer length scale (in the frame moving with the compression) is physical or not.

We mention that the rapid distortion theory (RDT) \citep{durbin2010,savill1987,hunt1990} solution to Eq.~(\ref{eq:moving_momentum_modeling}), with the viscous dissipation term included, can also give reasonable results when $\beta = 5/2$, the compression is fast, and the initial Reynolds number is not that high, as in the simulations presented here. In this case, the viscous RDT solution captures the initial growth well, and the decay well, but overshoots at the peak of the TKE. This overshoot decreases the more rapid the compression. Initially, the linear compressive forcing dominates the solution for a rapid compression. During the sudden dissipation, the linear viscosity term dominates the solution. Apparently, for $\beta = 5/2$, a modest Reynolds number, and a rapid compression, there is only a brief window for nonlinear effects between these two linear regimes. Note that, by $\beta = 3/2$, the nonlinear effects are very important, since this case corresponds to regular decaying NS turbulence in the rescaled equation, Eq.~(\ref{eq:scaled_momentum_modeling}). 

While we have discussed, in Sec.~\ref{sec:energy_partition_modeling}, the coupling of dissipated TKE to the temperature evolution, this treatment of the feedback only occurs in that section. In particular, the comparison to simulations is done for fixed $\beta$, while in the feedback case $\beta$ will effectively change as the rate of temperature growth with compression can change. The model should still be useful in this case, but it may not be as accurate as in the fixed $\beta$ cases where the comparison is carried out. Future work should compare the model to simulations that include the feedback of dissipated TKE to temperature (or, a variable $\beta$ that mimics it).

Apart from improving the handling of $L_{\rm{outer}}$, as discussed above, the present model may be improved upon by a model that includes the varying viscosity at a more fundamental level than the technique used here --- a (non-rigorous) assumption of non-constant viscosity in the model of \citet{lohse1994}. 

\textbf{SUMMARY} ---
We have presented a simple model for the TKE behavior of compressing subsonic turbulence, which agrees with simulations over a range of viscosity dependencies on compression. In this regard, it represents a substantial improvement over previously existing models. It is hoped this model will be useful for evaluating the prospects of preventing or exploiting plasma turbulence in plasma compression experiments.

\begin{acknowledgments}
This work was supported by NNSA 67350-9960 (Prime $\#$ DOE DE-NA0001836) and by NSF Contract No. PHY-1506122.
\end{acknowledgments}

\end{document}